 \documentclass{emulateapj}
\newcommand{\rxte}{{\it RXTE}}
\newcommand{\xte}{{\it RXTE}}
\newcommand{\cxo}{{\it CXO}}
\newcommand{\src}{1A~1744$-$361}

\def\lapp{\ifmmode\stackrel{<}{_{\sim}}\else$\stackrel{<}{_{\sim}}$\fi}
\def\gapp{\ifmmode\stackrel{>}{_{\sim}}\else$\stackrel{>}{_{\sim}}$\fi}
\def\nuddd{\ifmmode\stackrel{\bf \,{\ldots} }{\textstyle
    \nu}\else$\stackrel{\,{\ldots} }{\textstyle \nu}$\fi}
\def\nudddd{\ifmmode\stackrel{\bf \,{\ldots} .}{\textstyle
    \nu}\else$\stackrel{\,{\ldots} .}{\textstyle \nu}$\fi}
\shorttitle{An \ion{Fe}{26} Absorption Line in the Persistent Emission of \src}
\shortauthors{Gavriil et al.}
\begin{document}
\title{An  \ion{Fe}{26} Absorption Line in the Persistent
Spectrum of the Dipping Low Mass X-ray Binary  \src}
\author{Fotis P. Gavriil\altaffilmark{1, }\altaffilmark{2},  Tod
        E. Strohmayer\altaffilmark{1}, Sudip Bhattacharyya \altaffilmark{3}}
\altaffiltext{1}{NASA Goddard Space Flight Center, Astrophysics
                 Science Division, Code 662, Greenbelt, Maryland,
                 20771, USA}

\altaffiltext{2}{CRESST; University of Maryland Baltimore County,
                 Baltimore, Maryland, 21250, USA}

\altaffiltext{3}{Department of Astronomy and Astrophysics, Tata
                 Institute of Fundamental Research, Mumbai 400005,
                 India}

\begin{abstract}
We report on \textit{Chandra X-ray Observatory} (\cxo) High-Energy
Transmission Grating (HETG) spectra of the dipping Low Mass X-ray
Binary (LMXB) \src\ during its July 2008 outburst. We find that its
persistent emission is well modeled by a blackbody ($kT\sim1.0$~keV) plus
power-law ($\Gamma\sim1.7$) with an absorption edge at 7.6~keV. In the
residuals of the combined spectrum we find a significant absorption
line at 6.961$\pm$0.002~keV, consistent with the \ion{Fe}{26}
(hydrogen-like Fe) 2 - 1 transition. 
We place an upper limit on the   velocity of a redshifted flow  of 
$v < 221$~km~s$^{-1}$.
We find an equivalent width for
the line of $27^{+2}_{-3}$~eV, from which we determine a column
density of $7\pm 1 \times 10^{17}$~cm$^{-2}$ via a curve-of-growth
analysis.  Using XSTAR simulations, we place a lower limit on the
ionization parameter of $> 10^{3.6}$~erg~cm~s$^{-1}$. The properties
of this line are consistent with those observed in other dipping
LMXBs.  Using \textit{Rossi X-ray Timing Explorer} (\xte) data
accumulated during this latest outburst we present an updated
color-color diagram which clearly shows that \src\ is an ``atoll''
source. Finally, using additional dips found in the \xte\ and \cxo\
data we provide an updated orbital period estimate of
52$\pm$5~minutes.
\end{abstract}


\section{Introduction}
\src\ is a neutron star Low Mass X-ray Binary (LMXB) discovered by the
\textit{Ariel V} satellite \citep{dbi+76,ces+77}.  This source is a
transient LMXB, and several outbursts have been observed with a number
of missions, most notably the \textit{Rossi X-ray Timing Explorer}
(\xte).  In 2001 \citet{eac+01} discovered a burst from the direction
of \src, however, it could not be unambiguously identified as a
thermonuclear burst.  The first thermonuclear (type I) burst from
\src\ was discovered by \citet{bsms06} using \xte.  This type I X-ray
burst exhibited a 530-Hz burst oscillation which provided a
measurement of the neutron star's spin frequency. The
burst also provided an upper limit of $d < 9$~kpc, under the
assumption that its maximum luminosity could not exceed the Eddington
luminosity for a 1.4-$M_\sun$ neutron star \citep{bsms06}.

This source also shows regular incidents of intensity ``dips'' in its
X-ray emission.  According to \citet{fkl87}, such dips are produced by
obscuring material associated with a structured accretion disk. Dips
would only be visible for inclination angles 60$^\circ$$\lapp i
\lapp$80$^\circ$ \citep{fkl87}.  In principle dips could occur every
orbital cycle, however, because of variations in the obscuring
material in the disk, the presence of dips often varies from cycle to
cycle and from source to source.  The dips observed in \src, as is the
case for other dipping LMXBs, exhibit complex structure
\citep[see][Fig.~4]{bsms06}. By measuring the spacing between two
closely spaced dip episodes observed by \xte, \citet{bsms06} estimated
the orbital period of \src\ to be $\sim$97 minutes, however these dip
episodes were separated by a data gap, thus not precluding shorter
orbital periods.

Narrow absorption features have been seen in the persistent emission
of many ``dipping'' LMXBs.  For example, XB~1916$-$053~\citep{bpb+04,
idl+06}, X~1624$-$490 \citep{pob+02}, 4U~1323$-$62~\citep{bmd+05}.  By
far, the most prominent features are the \ion{Fe}{25} (He-like) and
\ion{Fe}{26} (H-like) lines. The ubiquity of narrow features in
dipping LMXBs suggests that they are a property of all LMXBs, but are
seen predominately in dipping LMXBs because their particular geometry
is optimal for viewing them \citep{dpb+06, bmd+05}.  These features
allow one to probe the structure, dynamics, and evolution of the
material surrounding the neutron star.  Thus, the spectral properties
of \src\ are of great interest.  Using \xte\ data, \citet{bssm06}
found that the persistent spectrum of this source is well modeled by a
Comptonized blackbody model \citep{bssm06}, and they also found
evidence of a broad ($\sim$0.6~keV) iron emission feature at
$\sim$6~keV and an iron absorption edge at $\sim$8~keV. \xte\ is not
sensitive to narrow spectral features given its coarse spectral
resolution and large FOV, thus, during the July 2008 outburst of this
source we triggered \textit{Chandra X-ray Observatory} (\cxo)
Target-of-Opportunity observations for this purpose.

In \S~\ref{sec:analysis} we present our analysis of the \cxo\
(\S~\ref{sec:hetg}) and \xte\ (\S~\ref{sec:xte}) data from \src. In
\S~\ref{sec:spec} we present high resolution spectra of the source in
which we find a \ion{Fe}{26} absorption feature.  In \S~\ref{sec:C-C
Diagram} we present an updated Color-Color diagram, and in
\S~\ref{sec:dips} we describe how we used the new observations of dips
to constrain the orbital period of the source. In
\S~\ref{sec:discussion} we discuss our results.


\section{Analysis}
\label{sec:analysis}

\subsection{\cxo\  Observations}
\label{sec:hetg}

During \src's last outburst in July 2008, we triggered three \cxo\ ToO
Observations. These observations are summarized in Table~\ref{ta:cxo
obs}. In order to search for narrow absorption features such as those
seen in other dipping LMXBs the data were taken with the High-Energy
Transmission Grating \citep[HETG][]{cdd+05} aboard \cxo. The HETG
consists of two arms, the High Energy Grating (HEG) and the Medium
Energy Grating (MEG). The HEG is sensitive to photons in the
0.8$-$10~keV band with a resolving power of $E/\Delta E$$\sim$1000 at
1~keV, and the MEG is sensitive to photons in the 0.4$-$5.0~keV band
with a resolving power of 660 at 0.8~keV. The gratings were operated
with ACIS which enables separation of the individual diffraction
orders, $m=\pm1,\pm2,\pm3$, and a nondispersed zeroth order image is
also obtained.  We inspected all dispersed orders, however, for our
analysis we concentrated on the $|m|=1$ data because its effective
area is more than an order of magnitude greater than the other orders
for both the MEG and the HEG.  Data in the different orders can be
combined, however they have different spectral and spatial responses
and thus we avoided doing so, lest we smear out any narrow features.

\begin{deluxetable*}{lllll}
\tablecolumns{5}
\tablewidth{2.0\columnwidth}
\tablecaption{\label{ta:cxo obs}}
\tablehead{ \colhead{Parameter\tablenotemark{a}} & \colhead{2008 July 5} & \colhead{2008 July 6} & \colhead{2008 July 7} & \colhead{Total} }
\startdata
Obs. ID  
& 9884 & 9885  & 9042  &  \nodata \\  
Exposure (ks) 
&   24.2   &    21.0   &  30.4     &         75.6  \\
Count Rate (counts s$^{-1}$) 
&  4.9$\pm$0.02  &  4.86$\pm$0.02      &  5.02$\pm$0.02     &   4.927$\pm$0.009 \\
\cutinhead{Continuum Model}
$N_H$\tablenotemark{b} ($\times 10^{22}$~cm$^{-2})$                                                        
& 0.40$^{+0.03}_{-0.03}$  & 0.35$^{+0.03}_{-0.03}$ &   0.44$^{+0.03}_{-0.03}$  & 0.414$^{+0.006}_{-0.015}$\\
$E_{\mathrm{Edge}}$\tablenotemark{c} (keV)                                                                 
& 7.56$^{+0.08}_{-0.05}$   & 7.73$^{+0.01}_{-0.01}$ &  7.699$^{+0.013}_{-0.008}$  &  7.645$^{+0.006}_{-0.015}$\\
$\tau_{\mathrm{Edge}}$\tablenotemark{d}                                                                    
& 1.8$^{+0.3}_{-0.2}$  & 3.5$^{+1.3}_{-0.7}$ & 3.5$^{+1.3}_{-0.6}$  & 1.17$^{+0.10}_{-0.09}$\\
$kT_\mathrm{bb}$\tablenotemark{e} (keV)                                                                    
& 0.98$^{+0.08}_{-0.08}$ & 1.05 $^{+0.11}_{-0.08}$ &  1.09 $^{+0.07}_{-0.06}$ & 1.06$^{+0.02}_{-0.03}$\\
$R_{\mathrm{bb}}$\tablenotemark{f}  (km)                            
& 4.8$^{+0.8}_{-0.7}$  & 4.4$^{+0.7}_{-0.6}$ & 4.4$^{+0.4}_{-0.3}$& 4.3$^{+0.2}_{-0.2}$\\
$L_{\mathrm{bol}}$\tablenotemark{g}  ($\times 10^{36}$~erg~s$^{-1}$) 
& 2.6$^{+0.3}_{-0.3}$    & 3.0$^{+0.6}_{-0.4}$ & 3.5$^{+0.5}_{-0.4}$ & 3.1$^{+0.2}_{-0.2}$\\
$\Gamma_{\mathrm{PL}}$\tablenotemark{h} 
& 1.64$^{+0.06}_{-0.06}$  &  1.51$^{+0.07}_{-0.05}$  & 1.80$^{+0.07}_{-0.06}$ &  1.68$^{+0.02}_{-0.02}$\\
Power-law Flux\tablenotemark{i} ($\times 10^{-9}$~erg~cm$^{-2}$~s$^{-1}$)                                  
&    1.26$^{+0.04}_{-0.04}$  &  1.39$^{+0.04}_{-0.06}$   & 1.14$^{+0.05}_{-0.06}$   & 1.25$^{+0.03}_{-0.02}$ \\
Total Flux\tablenotemark{j}  ($\times 10^{-9}$~erg~cm$^{-2}$~s$^{-1}$)                                     
&   1.399         & 1.368 & 1.241  & 1.367 \\
$\chi^2_{\mathrm{DoF}}$ [DoF] 
& 0.70 [1490] & 0.69 [1333] & 0.77 [1883] & 0.83 [3407]\\
\cutinhead{\ion{Fe}{26} line Properties}
$E_i$\tablenotemark{k} (keV)  
& 6.97$\pm$0.03 & \nodata  & 6.97$\pm$0.23 & 6.961$\pm$0.002\\
$\Delta E$\tablenotemark{l} (eV) 
& $<$50 &    \nodata     & $<229$ & 15.5$\pm$2.4\\
$W_E$\tablenotemark{m} (eV) 
& 33$^{+18}_{-15}$ & \nodata  & $<$40 & 27$^{+2}_{-3}$\\
FWHM\tablenotemark{n} (km s$^{-1}$) 
& $<$23192 & \nodata & $< 5065$ & 1613$\pm$21 \\
$N$\tablenotemark{o}  ($\times 10^{17}$~cm$^{-2}$) 
& 10$^{+12}_{-6}$ & \nodata & $<$13 & 7.4$^{+1.1}_{-1.0}$ \\
\enddata
\tablenotetext{a}{All errors represent 1-$\sigma$ uncertainties.}
\tablenotetext{b}{Hydrogen column density.}
\tablenotetext{c}{Absorption edge energy threshold.}
\tablenotetext{d}{Maximum depth of absorption edge at energy threshold.}
\tablenotetext{e}{Blackbody temperature.}
\tablenotetext{f}{Blackbody radius  as determined from the normalization of 
the blackbody component and  assuming a distance of $d$$=$9~kpc (the upper 
limit on the distance to the source).}
\tablenotetext{g}{Bolometric luminosity as determined from the normalization 
of the blackbody component and  assuming a distance of $d$$=$9~kpc (the upper 
limit on the distance to the source).}
\tablenotetext{h}{Index of power-law component.}
\tablenotetext{i}{Unabsorbed 2$-$10~keV flux of the power-law component.}
\tablenotetext{j}{Unabsorbed 2$-$10~keV flux of the combined model.}
\tablenotetext{k}{Line energy, see Eq.~\ref{eq:phi G}.}
\tablenotetext{l}{Line width, see Eq.~\ref{eq:phi G}.}
\tablenotetext{m}{Equivalent width of the line.}
\tablenotetext{n}{FWHM of the line in units of velocity.}
\tablenotetext{o}{Column density of the ion as determined by a COG analysis, 
see \S~\ref{sec:cog} for details.}

\end{deluxetable*}

\subsubsection{High Resolution Spectroscopy}
\label{sec:spec}

For each HETG arm and dispersed order we extracted a source and
background spectrum starting from the level 2 events file. We then
created a response matrix and ancillary response file for each arm and
order following the standard \cxo\ analysis threads. We then grouped
the spectra so that there were no less than 60 counts per bin after
background subtraction. We used
\texttt{XSPEC}\footnote{\url{http://xspec.gsfc.nasa.gov}} v12.5.1 for
subsequent spectral modeling. We fit the MEG data to a
photoelectrically absorbed blackbody and power-law model.  The MEG
data was well fit by this model, and there were no significant
deviations.  We then analyzed the HEG data and fit the same spectral
model with an absorption edge at $\sim$7.6~keV, because of the HEG's
higher response to photons above 5~keV. For the HEG we ignored bins
above 8~keV because of their diminished statistics. A
photoelectrically absorbed blackbody+power-law+edge model fits well,
see Table~\ref{ta:cxo obs}, however, the residuals showed a clear
deviation from the model at 6.96~keV in the 5 July 2008
(Fig.~\ref{fig:multi spec} panel 1A and 1B) and 7 July 2008
(Fig.~\ref{fig:multi spec} 2A and 2B) observations. There were no
comparable deviations in the 6 July 2008 observation, however, a hint
of the feature could be seen at 6.96~keV. We noticed that the feature
was primarily in the $m=-1$ and not the $m=1$ spectra, in principle
the two should have comparable spectra, since they have comparable
responses and effective areas in that energy range, however, dithering
could affect the spectrum and smear out the feature, which is what we
suspect happened to the order $m=1$ spectrum. We do not believe the
feature is an instrumental artifact because then it would be seen in
all $m=-1$ HETG observation of \src.

\begin{figure}
\plotone{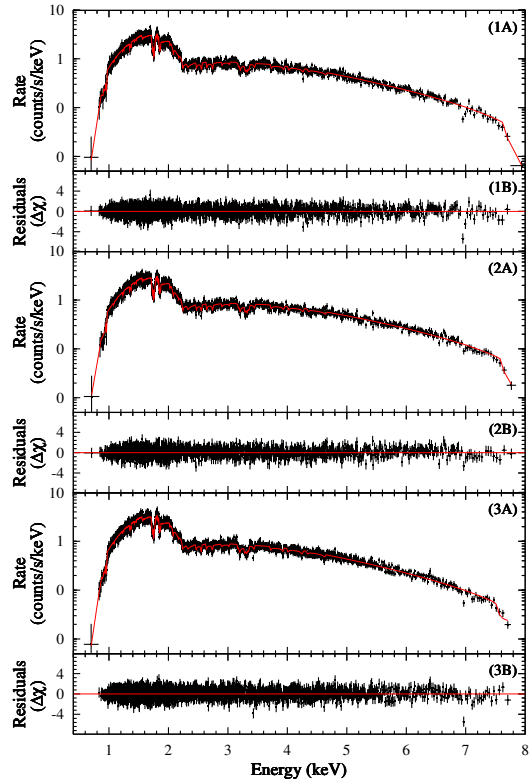}
\caption{HEG order $m=-1$ spectra of \src. Panels 1A, 2A, and 3A
corresponds to the 2008 July 5, July 6, and July 7 observation,
respectively. The red curve represents the best fit spectral model,
see Table~\ref{ta:cxo obs} for details.  Panels 1B, 2B, and 3B display
the residuals, in terms of $\sigma$ from the model, after fitting for
the best-fit spectral model.  Notice the highly significant deviations
from the model at 6.961~keV in Panels 1B and 3B \label{fig:multi
spec}.}
\end{figure}

\begin{figure*}
\plotone{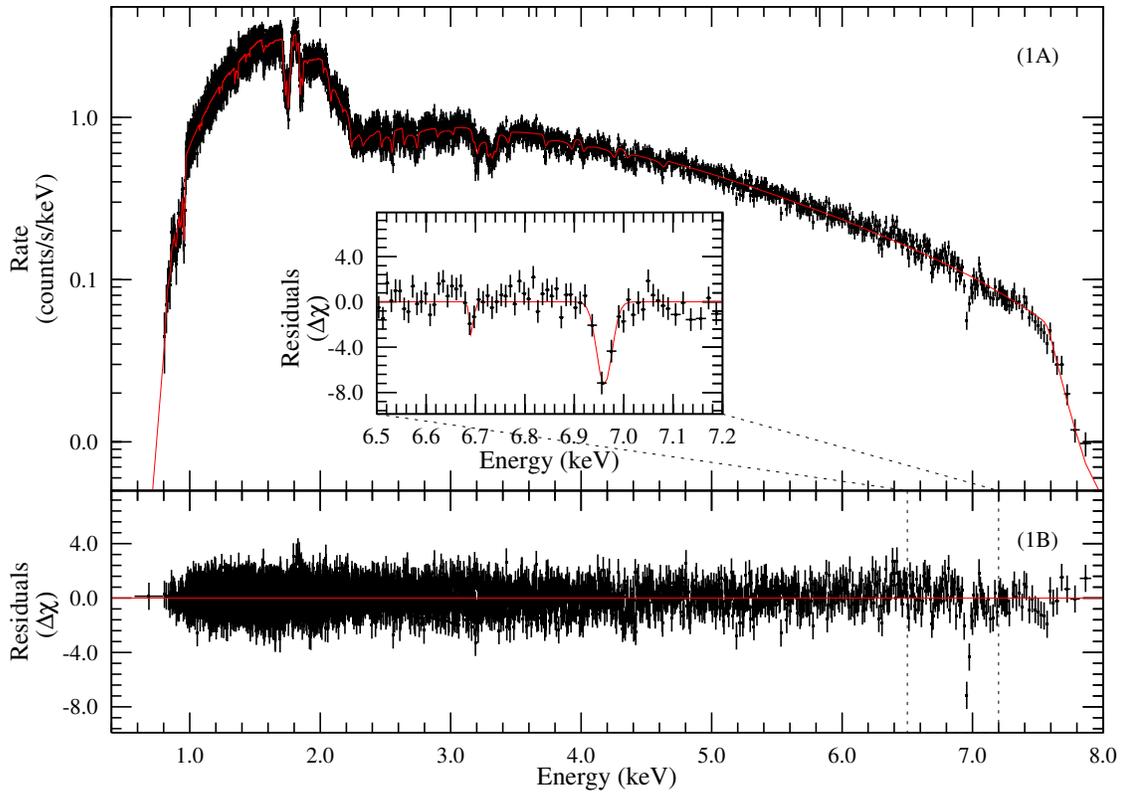}
\caption{The sum of the three HEG $m=-1$ spectra of \src\ displayed in
Fig~\ref{fig:multi spec}. The red curve is the best fit model
(photoelectrically absorbed blackbody plus power law times an edge,
see text for details). 1B: The residuals, in terms of $\sigma$ from
the model, after subtracting the best-fit model displayed in Panel
1B. Notice the highly significant deviations from the model at
6.961~keV corresponding to the \ion{Fe}{26} transition. Inset: An
expanded view of the residuals around the \ion{Fe}{26} feature.  The
red curve is the best fit Gaussian line profile which has a centroid
at $E_0 =6.961\pm0.002$~keV with a width of $\Delta E=15.5\pm2.3$~eV,
see text for details. The \ion{Fe}{25} is not statistically
significant, however we fit for it in order to place an upper limit on
its column density. \label{fig:total spec}}
\end{figure*}

In order to increase the signal-to-noise ratio of the feature we combined
all three observations and their ancillary response files. The
combined spectrum is displayed in Fig.~\ref{fig:total spec}, and the
6.96~keV feature can be clearly seen (Fig.~\ref{fig:total spec},
inset).  In the combined spectrum the line is clearly resolved.
Assuming that the line is broadened because of the thermal Doppler
effects, then the line profile close to the peak should resemble a
Gaussian:
\begin{equation}
\phi_G(E,kT) = \frac{1}{ \sqrt{2 \pi} \Delta E } 
  \exp{\left[{-\frac{\left(E-E_i\right)^2}{2 \Delta E^2 }
  }\right]},
\label{eq:phi G}
\end{equation}
where $E_i$ is the line energy of the ion, and the line width, $\Delta
E$, is given by,
\begin{equation}
{\Delta E} = E_i \left({kT}/{m_i c^2}\right)^{1/2},
\label{eq:kT}
\end{equation}
here, $m_i$ is the ion mass and $kT$ is the kinetic temperature
\citep[see][]{rl79}.  Fitting Eq.~\ref{eq:phi G} times a normalization
to the line profile we obtain a line energy of $E_i=6.961\pm0.002$~keV
and a standard deviation of $\Delta E=15.5\pm2.3$~eV.  The theoretical
properties of the \ion{Fe}{25} (He-like) and the \ion{Fe}{26} (H-like)
lines, the most prevalent lines in dipping LMXB spectra, are listed
in Table~\ref{ta:ions}. These values were obtained from the online
line finding
list\footnote{\url{http://heasarc.gsfc.nasa.gov/docs/software/xstar/xstar.html}}
provided by the \texttt{XSTAR} emission modeling package \citep{kb01}.
The peak energy is consistent with the \ion{Fe}{26} $n = 2 - 1$
transition, and according to Eq.~\ref{eq:kT}, our measured line width
corresponds to a kinetic temperature of $kT=254\pm2$~keV.  
Using the difference between the observed, $E_i$, and
rest energy, $E_0$, of the line (see Table~\ref{ta:ions}, we can determine the velocity of the
absorbing material via $(E_i-E_0)/E_i = v/c$, for which we find $v <
221$~km~s$^{-1}$. We can only place an upper limit of the flow
velocity because the wavelength difference we measure ($\Delta \lambda
= 0.001~$\AA) is less than the absolute wavelength accuracy ($\Delta
\lambda = 0.006$~\AA) of the HEG, see the Chandra X-ray Center's
Proposer's Observatory
Guide\footnote{\url{http://cxc.harvard.edu/proposer/POG/pog\_pdf.html}}.

\begin{deluxetable}{lccc}
\tablecolumns{4}
\tablewidth{\columnwidth}
\tablecaption{Theoretical Properties\tablenotemark{a} of the \ion{Fe}{25} (He-like) and \ion{Fe}{26} (H-like) line. \label{ta:ions}}
\tablehead{ \colhead{Ion} &   \colhead{$E_0$\tablenotemark{b} (keV)} & \colhead{$f$\tablenotemark{c}} & \colhead{$A_{ul}$\tablenotemark{d} ($\times$10$^{14}$~s$^{-1}$)}  }
\startdata
\ion{Fe}{25}   & 6.70010  & 0.775  & 5.033  \\
\ion{Fe}{26}   & 6.96614  & 0.408  & 2.863  
\enddata
\tablenotetext{a}{All parameters were provided by the \texttt{XSTAR} online line finding list \url{http://heasarc.gsfc.nasa.gov/docs/software/xstar/xstar.html}.}
\tablenotetext{b}{Energy of transition.}
\tablenotetext{c}{Oscillator strength.}
\tablenotetext{d}{Einstein coefficient.}
\end{deluxetable}

\subsubsection{Curve-of-Growth Analysis}
\label{sec:cog}
A curve-of-growth (COG) analysis allows one to relate the equivalent
width. $W_E$, of a spectral line to the column density, $N$, of the
absorbing ion.  \texttt{XSPEC} provides a function which computes the
equivalent width of a line, for which we obtain
$W_E=27^{+2}_{-3}$~eV. To verify our \texttt{XSPEC}-measured
equivalent width, we also computed the equivalent width analytically,
by assuming that in the spectral region of interest the continuum is
dominated by the power-law component.
Our analytically determined equivalent width, 29.8~eV, was consistent
with the value found using \texttt{XSPEC}.

\begin{figure}
\plotone{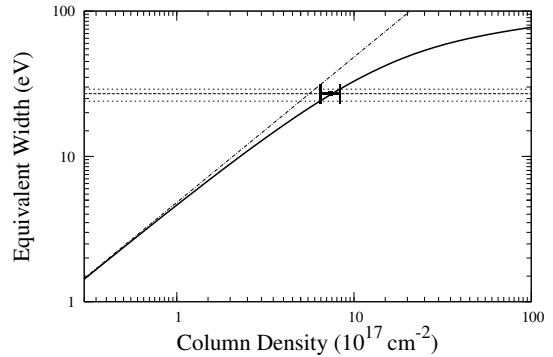}
\caption{Curve-of-Growth (COG) for the \ion{Fe}{26} absorption line
observed in the persistent emission of \src: Equivalent width versus
column density for a kinematic temperature of $kT=254$~keV. The error
on the COG due to the error on the temperature would be smaller than
the thickness of the line. The horizontal dashed line indicates the
position of our measured equivalent width, and the dotted lines are
the upper and lower bounds on the equivalent width. The dashed-dotted line
represents the extrapolation of the ``linear regime'' of the COG
\label{fig:COG}.}
\end{figure}

In order to calculate the column density we computed theoretical COGs
 following \citet{ked+00} and \citet{rl79}. The equivalent width,
 $W_E$, of the line is related to the optical depth, $\tau$ via
\begin{equation}
W_{E} = \int^{\infty}_0 (1 - e^{-\tau})dE,
\label{eq:width}
  \end{equation}
here $\tau = N s \phi$, where $N$ is the column density, $\phi$ is
the line profile shape, and
\begin{equation}
s =  {\pi f e^2}/{m_i c}.
\label{eq:s}
\end{equation}
where $f$ is the oscillator strength (see Table~\ref{ta:ions}), and $m_e$ and $e$ is the
electron mass and charge, respectively. Note, there is also a temperature
dependent term in Eq.~\ref{eq:s} which we ignored because
\citet{ked+00} showed that the contribution of this term is
negligible.  The overall line profile, $\phi(E,kT)$ is found by
convolving the line profile due to thermal broadening ($\phi_G$,
Eq.~\ref{eq:phi G}), and the line profile due to collisional
broadening ($\phi_L$), which has a Lorentzian profile, i.e,
\begin{equation}
  \phi_L(E) = \frac{h\gamma}{4\pi^2(E-E_0)^2 +(h\gamma/2)^2 },
\end{equation}
where $\gamma$ is the sum of the Einstein coefficients, $A_{ul}$,  over all lower energy states (see Table~\ref{ta:ions}).  Using the kinematic temperature found above, we then
numerically integrated Equation~\ref{eq:width} to obtain our COG which
is displayed in Figure~\ref{fig:COG}. Using the error on the dynamic
temperature and the equivalent width we can obtain a confidence region
for the \ion{Fe}{26} column density. We find $N_{\mathrm{Fe\;{XXVI}}}
= 7.4_{-1.0}^{+1.1}\times 10^{17}$~cm$^{-2}$ for \ion{Fe}{26}.

\subsubsection{Ionization Parameter}
\label{sec:xi}
\begin{figure}
\plotone{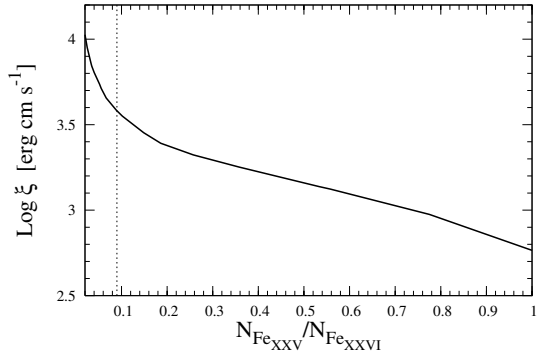}
\caption{Ionization parameter, $\xi$, as a function of the relative
abundance of \ion{Fe}{25} to \ion{Fe}{26} as determined by our
\texttt{XSTAR} simulations, see \S~\ref{sec:xi} for details. The
vertical dotted lines represents our upper limit on the relative
abundance of \ion{Fe}{25} to \ion{Fe}{26} which places a lower limit
on the ionization parameter of $\xi > 10^{3.6}$~erg~cm~s$^{-1}$
\label{fig:xi}.}
\end{figure}

Although the presence of the \ion{Fe}{26} feature is clear, there was
no discernible \ion{Fe}{25} feature.  However, an upper limit on the
column density of the \ion{Fe}{25} can constrain the ionization
parameter
\begin{equation}
 \xi = \frac{L}{n_e r^2},
\end{equation}
where $L$ is the luminosity of the source, $r$ is the distance from
the central object to the absorber, and $n_e$ is the electron number
density. The electron number density, is related to the electron
column density, $N_e$, via $n_e = N_e /\Delta r$, where $\Delta r$ is
the thickness of the absorbing slab.  In order to measure the
equivalent width of the \ion{Fe}{25} transition, we fit for such a
line using our combined spectrum, see Fig.~\ref{fig:total spec}
(inset). Then, using \texttt{XSPEC}, we measured the equivalent width
in the same manner as for \ion{Fe}{26}. We then generated a COG for
\ion{Fe}{25} using the corresponding oscillator strength and Einstein
coefficient (see Table~\ref{ta:ions}). From our COG analysis
we place an upper limit on the column density of \ion{Fe}{25} of
$N_{\mathrm{Fe\;{XXV}}} < 6.3\times 10^{16}$~cm$^{-2}$.  To calculate
$\xi$ we ran simulations of an ionized absorber using
\texttt{XSTAR}. We assumed a spherical distribution of material and we
used the standard spectral shape of a power-law with a spectral index
of 2. We created a grid of ionization parameters and relative
abundances using different initial choices for the electron density
and luminosity.  As an upper limit on the choice of luminosity we used
the bolometric luminosity as measured from the blackbody component, $L
< 3.5\times10^{36}$~erg~s$^{-1}$ (see Table~\ref{ta:cxo obs}).  Our
plot of $\xi$ as a function of the ratio of the column density of
\ion{Fe}{25} to \ion{Fe}{26} is shown in Fig.~\ref{fig:xi}.  Given
that the relative abundance of \ion{Fe}{25} to \ion{Fe}{26} is less
than 0.09, this places a lower limit on the ionization parameter of
$\xi > 10^{3.6}$~erg~cm~s$^{-1}$.

\subsection{\xte\ Observations}
\label{sec:xte}

The high time resolution data presented here were obtained from the
Proportional Counter Array (PCA) aboard \xte.  The PCA is made up of 5
independent proportional counter units (PCUs).  Each PCU is made up of
three Xenon/Methane layers and an uppermost propane veto layer. The
PCUs are sensitive to photons in the $\sim$2--60~keV band binned into
256 channels. The PCA and can time tag a photon to an accuracy of
$\sim$1~$\mu$s. \xte\ provides two standard data
modes. \texttt{Standard-1} mode lightcurves with 0.125-s temporal bins
and is summed over all spectral channels. \texttt{Standard-2} data
provides lightcurves with 16-s time bins and is grouped into 129
spectral channels.  As well as the standard modes, there are several
user selected modes available. The ones used here include
\texttt{GoodXenon} and \texttt{Event} mode. \texttt{GoodXenon} mode
provides the full temporal and spectral resolution.  \texttt{Event}
mode data provides high-time resolution but often at slightly lower
time resolution than \texttt{GoodXenon}, and the spectral channels are
grouped using different grouping schemes, but often preserving 64
spectral channels.

\subsubsection{Color-Color Diagram}
\label{sec:C-C Diagram}

We created a color-color diagram using the \texttt{Standard-2}
data. Each PCU has different energy response, and PCUs 0 and 1 have
lost their veto propane layer.  PCU 2 is the most reliable, thus in
our analysis we used only photons from the top Xenon layer of PCU 2.
For count rates below 20 counts/s we used 2048-s long intervals, for
count rates between 20 and 40 we took 512-s long intervals, and for
the highest count rates we used 256-s long time intervals.  For each
spectrum we generated a model background using the \texttt{FTOOL}
\texttt{pcabackest}. We defined our colors as follows: our soft color
is the ratio of counts in the 2.30$-$3.54~keV band to that in the
5.19$-$3.54~keV band, and our hard color is defined as the ratio of
the counts in the 5.19$-$8.52~keV band to that in the 8.52$-$17.74~keV
band. Our color-color diagram is displayed in Fig.~\ref{fig:C-C
diagram}. In \citet{bssm06} the authors reported that the source
exhibited atoll-like behavior, however a Z-source track could not be
unambiguously ruled out.  The additional \rxte\ data presented here
clearly display atoll source behavior.

\begin{figure*}
\plotone{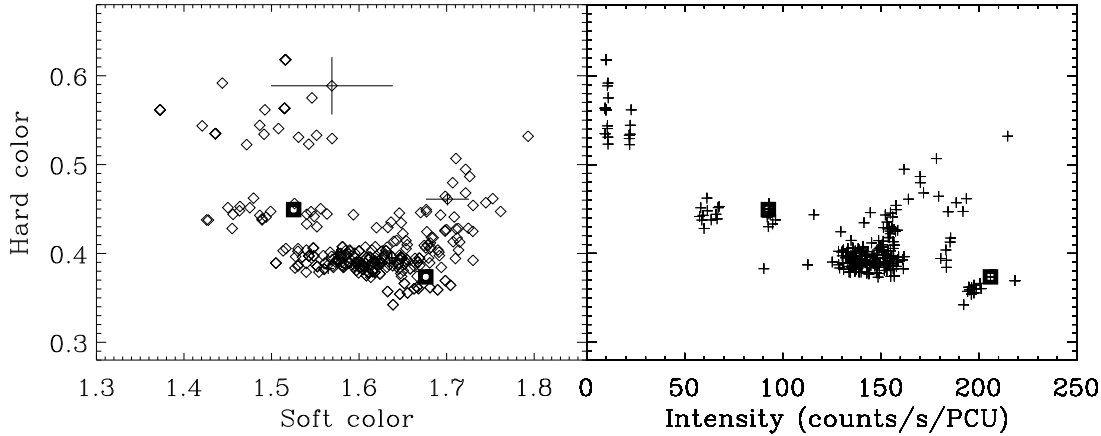}
\caption{Color-Color diagram of \src\ as observed by \xte. Left: Hard
color versus Soft color, see \S\ref{sec:C-C Diagram} for
details. Right: Hard color versus Intensity. From both panels notice
that \src\ exhibits atoll-like behavior. \label{fig:C-C diagram}}
\end{figure*}


\subsubsection{Timing Analysis}
\label{sec:timing}

\citet{bsms06} reported a tentative (2.3~$\sigma$ detection) of an
800-Hz quasi-periodic oscillation (QPO) for \src.  Using all the
available \xte\ event data of \src, we searched for low-frequency and
kHz QPOs. In our search we broke up our lightcurves into 250~s
intervals. We found no highly significant ($\gg$ 3-$\sigma$) QPOs.  The
most significant QPO we found was a low frequency QPO at 29~Hz QPO for
an observation on 2008 July 17 (Obs. ID 93155-01-06-03) at a
significance of 3.0~$\sigma$.
 
\begin{figure}
\plotone{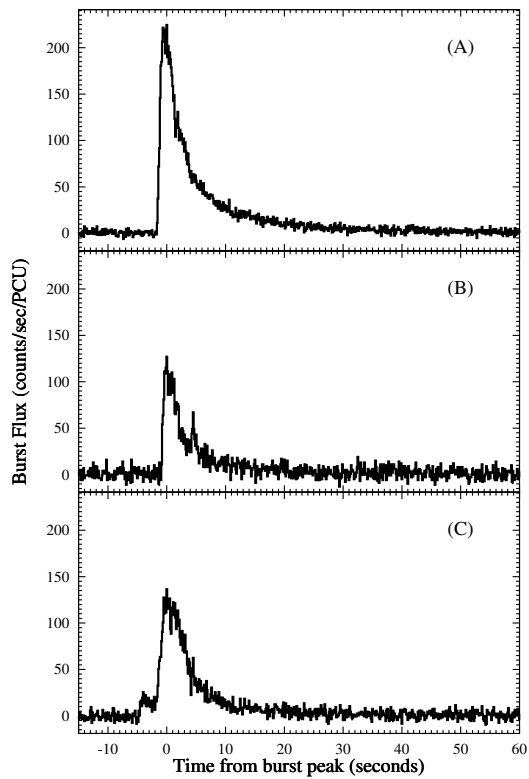}
\caption{The thermonuclear X-ray bursts observed from \src\ with
\xte. The lightcurves are background subtracted 0.125-S lightcurves
generated using the full PCA energy band. The bursts were observed on
2005 July 16 (A), 2008 June 24 (B), and 2008 July 25 (C).
\label{fig:bursts}}
\end{figure}

In the \xte\ archive we find a total of three thermonuclear bursts
from \src. These bursts are displayed in Figure~\ref{fig:bursts}. The
first burst occurred on 2005 July 16 (obs. ID 91050-05-01-00) and was
first reported by \citet{bsms06}.  This burst is the brightest 
observed from this source thus far (see Fig.~\ref{fig:bursts} Panel
A). Using its luminosity \citet{bsms06} placed an upper limit on the
distance to the source of $d<$9~kpc.  The second and third bursts
occurred on 2008 June 24 and 2008 July 25, respectively (see
Fig.~\ref{fig:bursts} Panels B and C).  \citet{bsms06} had found a
significant burst oscillation in the first burst. We searched the
other bursts for burst oscillations but found no comparable signals.

\subsection{Dip Timing  Analysis}
\label{sec:dips}
In an attempt to constrain the orbital period of \src,  we studied
the dips of this \src. There were 11 independent dip episodes in the  \xte\
archive, where a dip episode comprises of a single dip or a cluster of
closely spaced dips. A single short ($\sim$2~s) dip was also observed
in the 5 July 2008 \cxo\ observation. Using two closely spaced dip
episodes, \citet{bsms06} were able to estimate the orbital period of
\src\ as 97$\pm$22~minutes. However, these dip episodes were separated
by a data gap, thus not precluding a shorter period.  

\begin{figure}
\plotone{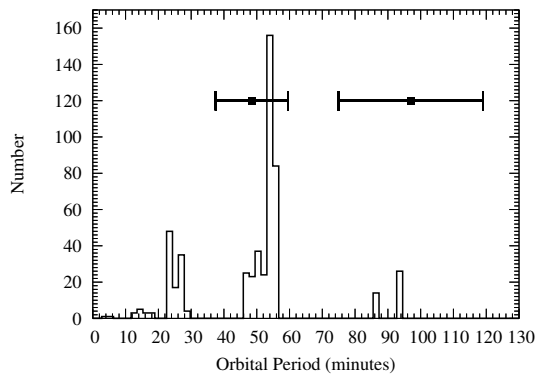}
\caption{Histogram of orbital periods that ensure that dips episodes
occur at the same orbital phase to within $\pm$0.2 cycles.  We stepped
through orbital periods between 3 and 120 minutes in steps of $10^{-4}$
minutes. The ordinate shows the number of periods, grouped into
1.8~minute bins, that satisfied the conditions we set out in
\S~\ref{sec:dips}.  The interval centered on 97~minutes shows the
interval derived by \citet{bsms06}. Our timing
analysis shows that the most likely orbital period is 52$\pm$5
minutes, which agrees within the uncertainties with half the orbital
period derived by \citet{bsms06} (the interval centered on 48.5).
\label{fig:orbital period}}
\end{figure}

In order to establish whether the period found by \citet{bsms06} is
consistent with the additional dips, we performed the following
analysis.  We generated an array of sine waves with periods between $P
= 3$ and $P=94$ minutes in steps of $10^{-4}$ minutes.  Assuming that
dips only occur in a particular range of orbital phase, we determined
which of the sine waves satisfy the following conditions for all dip
episodes: The center of all the dip episodes has to be within 0.2
cycles of the crest of the sine wave, and the 0.2~cycles around the
trough of the sine wave could not contain a dip-free interval. This
yielded 509 sine waves with orbital periods in our range that
satisfied the above conditions. We then grouped these remaining
permissible orbital periods into a histogram of bin size
1.8~minutes. The histogram is shown in Figure~\ref{fig:orbital
period}.  Notice that the orbital period found by \citet{bsms06} is
not ruled out, however, the observations seem to favor roughly half
that period.  The mean of the distribution displayed in
Figure~\ref{fig:orbital period} is 52~minutes and its standard
deviation is 5~minutes.  Thus we refine the orbital period estimate of
this source to 52$\pm$5 minutes.

\section{Discussion}
\label{sec:discussion}

Using three pointed \cxo\ HETG observations of \src\ during its latest
outburst we have discovered a significant 6.961~keV absorption feature
in its persistent spectrum.  The 6.961~keV absorption feature was
present in the first and third \cxo\ observation, and is well resolved
in the combined spectrum.  We have identified this feature as the
\ion{Fe}{26} (hydrogen-like Fe) 2 - 1 transition . We find no evidence
for lines due to other transitions.  This is the first time such a
narrow absorption feature has been observed in the spectrum of \src,
making this source another example of a dipping LMXB which shows
narrow absorption features.  Because of the ubiquity of these
features, it has been suggested that all LMXBs should exhibit them and
that they are predominately observed from dipping LMXBs because their
relatively edge on geometry facilitates such measurements
\citep{dpb+06, bmd+05}. Our discovery of such a feature in this
dipping LMXB definitely corroborates that hypothesis.

The feature we have detected is consistent with the wavelength of the
\ion{Fe}{26} $n = 2-1$ transition to the wavelength accuracy of the
detector.  We place an upper limit on the velocity of a redshifted
flow of $v < 221$~km~s$^{-1}$.  In the combined spectrum, the line is
well resolved, and we measure a line width of $\Delta
E=15.5\pm2.3$~eV. Under the assumption that the line is broadened due
to thermal effects, the measured line width corresponds to a dynamic
temperature $kT=254\pm2$~keV.  Using this temperature we performed a
curve-of-growth analysis for \ion{Fe}{26} and determined a column
density for \ion{Fe}{26} of $N_{\mathrm{Fe\;{XXVI}}} =
7.4_{-1.0}^{+1.1}\times 10^{17}$~cm$^{-2}$. This column density is
comparable to that found for other dipping LMXBs. For example, for
XB~1916$-$053 \citet{idl+06} found $N_{\mathrm{Fe\;{XXVI}}} =
6.6\times 10^{17}$~cm$^{-2}$. We note however, that they used the
linear approximation of the COG, i.e., for low column densities
Equation~\ref{eq:width} reduces to $W_E \approx (\pi f e^2/h m_i
c)N$. However from Fig.~\ref{fig:COG} our features appears to lie
beyond the linear regime.  Unlike XB~1916$-$053 we do not find
statistical evidence for a feature at the \ion{Fe}{25} energy,
indicative of the presence of a highly ionized absorber for \src.  Our
upper limit on the equivalent width on an underlying \ion{Fe}{25}
feature places a lower limit on the ionization parameter of $\xi >
10^{3.6}$~erg~cm~s$^{-1}$, as determined by our \texttt{XSTAR}
simulations. We can not state with certainty that \src\ is more highly
ionized than other dipping LMXBs where both the \ion{Fe}{25} and
\ion{Fe}{26} are present because our lower limit on $\xi$ is not above
the values found for those sources, in particular XB~1916$-$053 and
4U~1323$-$62 which had $\log \xi = 4.15$ and $\log \xi = 3.9$,
respectively \citep{idl+06, bmd+05}.

\src\ is a prolific dipper, we have identified a total of 11 dips
episodes in the \xte\ data and 1 in the \cxo\ data.
\citet{bsms06} determined an orbital period of 97$\pm$22~minutes from
two consecutive dip episodes, however a data gap made it uncertain
whether this was the true orbital period or a multiple thereof.  In
\S~\ref{sec:dips} we propose that if the dips occur within a
particular range of orbital phases, then the source's  orbital period is more
likely to be near half the value reported by \citet{bsms06},
specifically $P=52\pm5$~min.  Some of \src's dips, as observed by
\xte, are long and exhibit a great deal of structure, see Fig.~4 of
\citet{bsms06}, for example.  Unfortunately, in our \cxo\ observations
we observed a single short (2-s long) dip. With additional data from
imaging telescopes such as \cxo, it would be interesting to compare
the spectral properties of \src\ during dipping and non-dipping
intervals.  Variations in the properties of any observed narrow
absorption lines would provide information about the structure and
evolution of the material around the neutron star. Such an analysis
was possible for 4U~1323$-$62, for which \citet{bmd+05} found that the
ionization parameter of 4U~1323$-$62 varied between dipping and
non-dipping intervals.

In PCA data of \src\, \citet{bssm06} discovered a broad (0.6~keV) Fe
emission line at 6~keV. The fact that this feature was below 6.4~keV
prompted \citet{bssm06} to predict that there might be an absorption
feature at $\sim$7~keV that was unresolved due to \xte's course
spectral resolution. No broad emission feature were observed in the
individual or combined \cxo\ observations, however this could be due
to the smaller area of \cxo\ as compared to \xte. By combining
additional observations one may be able to observe both the Fe
emission feature and the \ion{Fe}{26} feature in \src. Indeed, such a
measurement was made for 4U~1323$-$62 by \citet{bmd+05}, using
\textit{XMM-Newton}. If future high spectral resolution observations
of \src\ reveal such an emission feature, and it is relativistically
broadened, then such measurements can potentially constrain the
compactness of the star.

\section{Conclusions}
\label{sec:conclusion}

We present \cxo\ HETG observations of \src\ during its latest
outburst, in which we have discovered a significant absorption feature
at 6.961$\pm$0.002~keV in its persistent emission.  We identify this
feature as a \ion{Fe}{26} line, analogous to those seen in other
dipping LMXBs. The width of the line was found to be $\Delta
E=15.5\pm2.4$~eV, which corresponds to a dynamic temperature of
$kT=254\pm2$~keV.  We place an upper limit on the velocity of a
redshifted flow of $v < 221$~km~s$^{-1}$.  The line had an equivalent
width of $W_E=27^{+2}_{-3}$~eV, for which we find a column density of
$N_{\mathrm{Fe_{XXVI}}} = 7\pm 1 \times 10^{17}$~cm$^{-2}$ via a COG
analysis.  We place an upper limit on the ionization parameter of $\xi
< 10^{3.6}$~erg~cm~s$^{-1}$.  All these values are consistent with the
\ion{Fe}{26} features seen in other dipping LMXBs.  Using additional
\xte\ data we generated an updated color-color diagram for the source,
which confirms the suggestion of \citet{bssm06} that this source is an
``atoll'' source.  We report two bursts in addition to the one
reported by \citet{bsms06}.  Performing a ``dip'' timing analysis, and
under the assumption that dips only occur at a particular range of
orbital phases, we find that $P=$52$\pm$5~minutes is the most likely
orbital period, nearly half the frequency found by \citet{bsms06}.
\src\ has demonstrated many of the properties of a canonical dipper,
however, because it not as well sampled as the others, many of its
parameters remain unconstrained.  Further observations with \xte\ and
\cxo\ during its next outburst are definitely warranted.

\acknowledgements We thank C.~B.~Markwardt for useful discussions and
for providing support for his numerical integration algorithm.  This
work has been supported by    NASA via an ADP grant and a
\textit{Chandra} Guest Observer grant, as well as by the National
Science Foundation (US NSF grant AST 0708424).  This research has made
use of data obtained through the High Energy Astrophysics Science
Archive Research Center Online Service, provided by the NASA/Goddard
Space Flight Center.


\end{document}